# THE CRITICAL DYNAMICS OF THE MODELS OF IRON-VANADIUM MAGNETIC SUPERLATTICE

Akai K. Murtazaev [1,2], Vadim A. Mutailamov[*] [1]

[1] Institute of Physics DSC RAS, 94 M.Yaragskii Str., Makhachkala, Russia, 367003.

[2] Daghestan State University, 43a M.Gajiev Str., Makhachkala, Russia, 367025.

[*] Corresponding author. E-mai address: vadim.mut@mail.ru

ABSTRACT

We report the results of a numerical investigation of static and dynamic critical behavior of the anisotropic easy-plane Heisenberg model which is used as a model for iron-vanadium magnetic superlattices. Models of iron-vanadium magnetic superlattice investigated with different values of intralayer and interlayer exchange interactions ratio. Numerical experiment technique, based on the dynamic finite-size scaling theory. Basic features of the time evolution of dynamic parameters for the researched systems are studied. The effect of interlayer exchange interaction value on a character of the critical dynamics is determined. It is confirmed the possibility of application of the Hamiltonian with strong easy-plane anisotropy for investigation both static and dynamic critical behavior of complex planar magnetic models.

Keywords: phase transitions; critical phenomena; finite-size scaling; critical dynamic; numerical experiment; anisotropic easy-plane Heisenberg model.

## 1. INTRODUCTION

The metallic magnetic superlattices consisting of alternating atomic layers of magnetic and nonmagnetic materials arouse great interest in the modern condensed-matter physics [1-3]. The possibility to control the basic properties of superlattices (magnetization, interlayer exchange interaction, magnetoresistance and other characteristics) via the external interaction allows creating structures with predetermined parameters, what makes these materials the unique objects for practical application and theoretical investigation [1-3]. Moreover, the magnetic superlattices provide an ideal facility for practical observance of the continuous crossover from three- to two-dimensional magnetization and inversely.

A situation with investigation of the critical properties of magnetic superlattices is rather complicated at present since the available results are contradictory [4, 5]. The experimental researches of such systems require the materials of very high quality. A production of high-quality samples and high-precision investigation of their critical properties is an extremely complex problem. Most likely this is the reason of few numbers of works, where the magnetic superlattices are studied near a phase transition point, and the reason for the contradictory results. Therefore, the methods of computational physics have been used to study the critical properties of magnetic superlattices in recent years. In works [6-8] the static critical behavior of Fe/V magnetic supperlattices is investigated, static critical exponents are calculated, and their dependence on ratio of intralayer and interlayer exchange interactions is studied. Methods of computational physics such as Monte Carlo and molecular dynamics methods have several advantages. Methods is based on a rigorous mathematical foundation. They allow you to control the error in the framework of the numerical experiment. In addition, they allow us to determine a degree of influence on the results of one or another parameter.

Great interest also causes the investigation of absolutely unstudied dynamic critical behavior of the magnetic superlattices. The investigation of the dynamic critical properties is one of actual questions of the modern statistical mechanics and phase transition physics [9-11]. To the present day the significant progress is achieved in this field due to the theoretical and experimental investigations mainly. Nevertheless, a development of a rigorous and consistent theory of dynamic critical phenomena based of microscopic Hamiltonians remains a central problems of modern theory of phase transitions and critical phenomena and still is far from solution [9, 12].

Rigorous theoretical investigation of the critical dynamics of spin systems based of microscopic Hamiltonians is an extremely difficult question even for the simple spin models; a situation with study of critical properties of magnetic superlattices is more complicated. Investigating the static critical properties of Fe/V superlattices it was established that critical exponents depend on a value of interlayer exchange interaction [6-8]. At the same time the scaling relations between critical exponents satisfy with very high accuracy. This situation does not fit into the framework of the modern theory of phase transitions and critical phenomena. Therefore, the investigation of critical dynamics is not only a considerable interest in itself, but it may become important for interpretation of difficulties encountered in the study of static critical phenomena.

The key characteristics of critical dynamics are the critical exponent $w$ for the relaxation time $\tau$ and the dynamic critical exponent $z$:

$$\tau \infty |t|^{-w}, \tau \infty \xi^z,$$

where

$$t = |T - T_c|/T_c, \; \xi = (T/T_c - 1)^{-\nu}.$$

In the mid-1990s, a new technique was proposed for calculating the dynamic critical exponent *z*, based on dynamic finite-size scaling theory [13] and a special procedure for determining the characteristic frequency $\omega_c$ [14-20].

In this paper we have studied a static and dynamic critical behavior of models of $Fe_2/V_{13}$ superlattices at two values of the interlayer and intralayer exchange interaction ratios.

2. MODEL

In [6-8] proposed a microscopic model of iron-vanadium superlattice $Fe_2/V_{13}$. Each iron atom has four nearest neighbors in the adjacent iron layer. The iron layers are shifted relative one another on half lattice constant along the *x* and *y* axes. The magnetic moments of iron atoms are ordered in *xy* plane. The scheme of iron-vanadium sublattice is shown in Fig.1.

An interaction between the nearest neighbors within layer has a ferromagnetic character and is determined by the exchange interaction parameter $J_\parallel$. The interlayer interaction $J_\perp$ between vanadium magnetic layers is transferred by the conduction electrons in the non-magnetic interlayer of vanadium (RKKI-interaction). In the real sublattices, its value and sign can change depending on a number of adsorbed hydrogen into the vanadium subsystem. Since an accurate dependence of the RKKI-interaction is unknown, usually, when carrying out the numerical investigations the whole range of interlayer interaction values from $J_\perp = -J_\parallel$ to $J_\perp = J_\parallel$ is studied.

As in the experiment the distance between magnetic layers is substantially larger than the interatomic distance, every atom interacts with the averaged moment of neighbor layers. A size of averaging region is a model parameter. Our investigations are made for limiting case, when every atom interacts with only one nearest atom from neighbor layer. The study of the static critical behavior of magnetic superlattices [6-8] showed that this approach describes a critical behavior of these models to the best advantage.

In [6-8] the static critical behavior of $Fe_2/V_{13}$ superlattices is investigated using a Hamiltonian of modified 3d *XY*-model:

$$H = -J_\parallel \frac{1}{2} \sum_{i,j} \left( S_i^x S_j^x + S_i^y S_j^y \right) - J_\perp \frac{1}{2} \sum_{i,k} \left( S_i^x S_k^x + S_i^y S_k^y \right), \quad (1)$$

where first sum takes into account the direct exchange interaction of each magnetic atom with nearest neighbors inside the layer, and second denotes the RKKI-interaction with atoms of neighboring layers through non-magnetic interlayer; $S_i^x$ and $S_i^y$ are *x* and *y* components of the spin localized in a site *i*. At numerical experiment a ratio $R = J_\perp/J_\parallel$ between interlayer exchange and intralayer exchange is a given parameter and can vary from $R = -1.0$ to $R = 1.0$ [6-8].

Hamiltonian (1) describes well the static critical behavior of models of iron-vanadium superlattices, but it can not be used for investigation of their dynamic critical behavior. As a rule, a method of molecular (spin) dynamics, based on a solution of spin motion equations in the local magnetic field, is used as numerical method for investigation of the spin dynamics of the magnetic systems [21]

$$\frac{\partial \vec{S}_i}{\partial t} = [\vec{S}_i \times \vec{h}_{loc}], \quad t = t'(g\mu_B/J\gamma), \quad |\vec{S}_i| = 1, \quad (2)$$

where $g$ is Landé factor; $\gamma$ is gyromagnetic ratio; $\vec{h}_{loc}^i$ is local magnetic field effecting on $\vec{S}_i$ and defined by the Hamiltonian. However, due to the mismatch of the spatial dimension of the spins in (1) and (2) directly using this method fails in the case of the model of iron-vanadium superlattice. Therefore, a special technique based on application of 3d Heisenberg model with strong anisotropy of the exchange constant in $z$ direction for studying similar systems was offered in [22]. According to this method we propose a model for investigation both static and dynamic critical behavior of iron-vanadium superlattice models. A Hamiltonian of this model can be written as:

$$H = -\frac{1}{2}\sum_{i,j}\left(J_\parallel^x S_i^x S_j^x + J_\parallel^y S_i^y S_j^y + J_\parallel^z S_i^z S_j^z\right) - \frac{1}{2}\sum_{i,k}\left(J_\perp^x S_i^x S_k^x + J_\perp^y S_i^y S_k^y + J_\perp^z S_i^z S_k^z\right), \quad (3)$$

$$J_\parallel^x = J_\parallel^y, J_\parallel^z = 0, \quad J_\perp^x = J_\perp^y, J_\perp^z = 0.$$

As can be seen from the equation, the Hamiltonian (3) is similar to the Hamiltonian of the 3D Heisenberg model. The difference is that the interaction of the spin components of $z$ in (3) is equal to zero. The lack of interaction of $z$ spin components leads to effect of strong anisotropy in plane $xy$ what points to the planar behavior of this model. In case of equality of the interlayer and intralayer interactions we can indicate the similarity of behavior between the studied model and classical 3D *XY*-model. This suggestion was confirmed in [22], where the authors showed that the static critical exponents of Heisenberg model with easy-plane anisotropy (obtained by numerical sumulation) coincide with values of similar exponents for 3D *XY*-model. But the critical temperature of these models differ from each other. Also we tested this method by investigating the critical dynamics of Heisenberg model with strong easy-plane anisotropy on a simple cubic lattice [23]. The obtained results showes that such approach describes well both static and dynamic critical behavior of planar magnetic models, including models of magnetic superlattices.

3. INVESTIGATION METHOD.

Our approach relies on dynamic scaling theory [13] and calculation of space- and time-displaced spin-pin correlation functions for spin components $k=x,y,z$

$$C^k(\vec{r}_{12},t) = \frac{\langle S^k_{\vec{r}_1}(t)S^k_{\vec{r}_2}(0)\rangle - \langle S^k_{\vec{r}_1}(t)\rangle\langle S^k_{\vec{r}_2}(0)\rangle}{\langle S^k_{\vec{r}_1}(0)S^k_{\vec{r}_1}(0)\rangle - \langle S^k_{\vec{r}_1}(0)\rangle\langle S^k_{\vec{r}_1}(0)\rangle}, \qquad (4)$$

where $\vec{r}_{12} = \vec{r}_1 - \vec{r}_2$, $S^k_{\vec{r}_1}(t)$ is a spin localized in a site $\vec{r}_1$ at instant time $t$, $S^k_{\vec{r}_2}(0)$ is a spin localized in a site $\vec{r}_2$ at the initial time ($t$=0), and angle brackets denote ensemble averaging.

The space-time Fourier transform of (4) defines the dynamic structure factor [9]

$$S^k(\vec{q},\omega) = \int d\vec{r} \int_{-\infty}^{+\infty} C^k(\vec{r},t)e^{-i(\vec{q}\vec{r}-\omega t)}dt, \qquad (5)$$

where $\vec{q}$ is a wave vector, and $\omega$ is a frequency. In the general case, the dynamic structure factor is proportional to the directly measurable neutron scattering function, being shifted by a constant frequency [9].

In accordance with the dynamic scaling hypothesis the characteristic frequency is the median frequency defined by the relation

$$\int_{-\omega_c(\vec{q},\xi)}^{+\omega_c(\vec{q},\xi)} S^k(\vec{q},\omega)d\omega = \frac{1}{2}\int_{-\infty}^{+\infty} S^k(\vec{q},\omega)d\omega. \qquad (6)$$

In the general case a characteristic frequency depends on wavevector $\vec{q}$ and correlation length $\xi$. In dynamic scaling theory [9], it is postulated that

$$\omega_c(\vec{q},\xi) = q^z \Omega(q\xi), \qquad (7)$$

where $z$ is a dynamic critical exponent, $\Omega$ is a an unknown homogeneous function of $q\xi$.

In the models considered here, expression (5) for a system of finite size $L$ simulated at the critical point over a finite time interval $t_{cutoff}$ is represented as [14, 16]

$$S^k(\vec{q},\omega) = \frac{1}{\sqrt{2\pi}} \sum_{\vec{r}_1,\vec{r}_2} e^{i\vec{q}(\vec{r}_1-\vec{r}_2)} \int_{-t_{cutoff}}^{+t_{cutoff}} e^{i\omega t} C^k(\vec{r}_1-\vec{r}_2,t)dt, \qquad (8)$$

and equation (7) becomes

$$\omega_c \sim L^{-z}\Omega'(qL), \qquad (9)$$

where $q=2\pi n/L$ ($n=\pm 1, \pm 1,...,L$). The sum in (8) corresponds to an integral over the space discretized into a lattice. Relation (9) is used to evaluate $z$ in practical calculations. The function $\Omega'$ is unknown, but it is known to depend only on the product of $q$ and $L$. Therefore, the dynamic critical exponent can be evaluated by keeping it constant while varying the lattice size.

In our study, we consider the case when the wave vector is directed only along the crystallographic axes (coordinate axes). Averaging over spins lying in alternating planes perpendicular to this axis was performed, and the averaged spins were used to calculate spin–spin correlation functions. For a simple cubic lattice spin planes coincide with the crystallographic planes. The geometry corresponding to the $Fe_2/V_{13}$ superlattice is more complicated.

The correlation functions were computed by using the system of differential equations of spin dynamics (2). Before solving system (2), the standard Metropolis algorithm [24] is executed to thermalize the system at the critical temperature.

4. RESULTS.

4.1. Static critical behavior.

The dynamic critical behavior of models of magnetic materials must be investigated at the phase transition point. For finding the critical temperatures of $Fe_2/V_{13}$ magnetic superlattices we preliminarily studied their static critical behavior. The systems with periodic boundary conditions, linear sizes from $L=8$ to $L=48$ containing from $N=512$ to $N=110592$ spins, and ratio of interlayer and intralayer exchange interactions $R=1.0$ and $R=0.7$ were investigated. Calculations were performed using a standard algorithm of the Monte Carlo method.

The initial non-equilibrium part of the Markov chain in the 70 000 Monte Carlo steps on spin, which is obviously larger than the relaxation time of the system are discarded in order to bring the spin system to equilibrium. In equilibrium, the average values of thermodynamic quantities are calculated. The length of the equilibrium part of the Markov chain ranged from 150 000 to 200 000 Monte Carlo steps on spin, depending on the linear dimensions of the system. In addition, to improve the statistics at each temperature for each linear dimension investigations were carried out at 10 different initial equilibrium spin configurations. Then the obtained results were averaged with each other. Critical temperatures were calculated using fourth-order Binder cummulants [25].

Our values of the critical exponents of the investigated models presented in Table 1. They are close to those obtained earlier in [6-8] for the same models, but using the Hamiltonian (1). When the interlayer and intralayer exchange interactions are equal ($R=1.0$) our data are also close to the theoretically predicted values for the classical 3D *XY*-model [26]. The distinction in critical temperatures is caused by difference in Hamiltonians of the researched models. The obtained values of static critical exponents of iron-vanadium superlattices confirms the planar nature of the critical behavior of these models.

Fig. 2 presents a dependence of instantaneous values of magnetization vector components (the order parameter) on the Monte-Carlo step number per spin. The data are presented for the system with spin number $N=2744$ and with the ratio of exchange interaction $R=1.0$ being in an equilibrium state near $T_c$. It is clear that in process of "time" *z*-component of the magnetization vector remains almost equal to zero, what also defines a behavior of $Fe_2/V_{13}$ superlattice as a model of planar magnetic.

4.2. Dynamic critical behavior.

When studying the dynamic critical behavior of iron-vanadium superlattices the ivestigated system initially was brought into the sate of thermodynamic equilibrium in the critical point by the standard algorithm of Monte-Carlo method. As a result a two-component magnetization vector $\vec{M}_{xy}$ in the *xy* plane took some arbitrary direction. To determine the longitudial and transversal components of the dynamic structure factor crystal lattice was reoriented in a space so that vector $\vec{M}_{xy}$ was directed along the *x* axis. As a result the *y*-component of this vector became zero and it modulus remained invariable. This state was accepted as an original configuration for the launching of the molecular dynamics method, in the course of which the equations of spin motions in local magnetic field (2) were measured by fourth-order Runge-Kutta method. Space–time spin correlation functions (4) were calculated according to results of equation solutions.

The systems with periodic boundary conditions with spin number from $N=512$ to $N=8000$ at ratio of exchange interactions of $R=1.0$ and $R=0.7$ were investigated. The length of the Markov chain in which the spin system was brought into a state of thermodynamic equilibrium was 70 000 Monte Carlo steps per spin. During the simulation of molecular dynamics total observation time of the spin system was up to $t_{cutoff}=120$ (in arbitrary units). Step numerical integration of differential equations of motion was taken equal to $\Delta t=0.01$. The number of averages in the calculation of the correlation functions (4) was $n=300$. For each system, with the number of spins N for each ratio R, the simulation process was carried out up to 10 times at different original equilibrium space configuration. The results were averaged between each other.

Fig. 3 presents a typical form of obtained space-time spin correlation functions. The data are presented for the system with $N=1728$ and ratio of exchange interactions $R=1.0$ being in equilibrium state at $T=T_c$. The correlations were examined for *x*-components of spins along the line *Ox*-axis. It is obvious that correlations decay both at spacing and in process of time. A dependence of correlation functions on the time at fixed value $|\vec{r}|=0$ for the systems with different spin number is shown in Fig. 4. It is noticeable the increase in relaxation time of the system at growth of the system linear sizes in accordance with dynamic finite-size scaling theory. In Fig. 5 there are shown similar dependences for the systems with $N=1728$ at two values of exchange interactions ratio. Our data failed to reveal the marked difference in a behavior of space-time spin correlation functions with a change in the value of interlayer exchange interaction.

The estimated values of correlation functions allowed us to calculate the dynamic structure factors for all studied systems. Since a spin system was oriented towards the magnetization vector after bringing the system into the thermodynamic equilibrium, we calculated the longitudinal $S_L(\vec{q},\omega)$ and transverse $S_T(\vec{q},\omega)$ dynamic structure factors. The longitudinal structure factor was estimated by the correlation functions values derived from *x*-components of spins, and the transverse structure factor was get by *y*-components. Our data showed that there is no difference in longitudinal and transverse components of the dynamic structure factor for the system with both values of *R*. Let us note that authors of [22] investigated the dynamic critical behavior of classical 3D *XY*-model also did not find a difference between these two components.

The abovementioned is illustrated in Fig.6, where the structure factor are presented in the dependence on frequency for system with spin number *N*=1728 at ratio of exchange interactions *R*=1.0. The wave vector is directed along *Ox* axis. The value *qL*, in accordance with (9), is taken equal to 2π. The same situation was observed at ratio *R*=0.7. When reducing the ratio *R* our data showed a little decrease in maximum of the structure factors for systems with the identical spin number and a slight change in their form. So, for example, in Fig. 7 are illustrated the longitudinal dynamic structure factors for the system with *N*=1728 at two values of exchange interaction ratio *R* and small values of frequency ω.

By the frequency dependences of dynamic structure factors we calculate the characteristic frequencies $\omega_c$ for all examined systems. Fig. 8 demonstrates the dependences of characteristic frequencies on the system linear sizes for two values of exchange interactions ratio *R* in a double logarithmic scale. On a diagram the characteristic frequencies estimated by the longitudinal dynamic structure factors are denoted by $\omega^L_c$, and $\omega^T_c$ means the frequencies received by the transverse ones. The wave vector $\vec{q}$ was selected in the line of *Ox* and *Oy* axes. As follows from the figure, when increasing the linear sizes the values of characteristic frequency decrease. And in accordance with the dynamic finite-size scaling law (9) this reduction in double logarithmic scale is a linear dependence. The results show that there is no significant difference in the values of the characteristic frequencies obtained from both the longitudinal and transverse structure factors in any direction of the wave vector.

The dependence of the characteristic frequency on the linear dimensions of the system allows using the expression (9) to determine the value of the dynamical critical exponent *z*. Unfortunately, our results were not sufficient to calculate the exact value of this exponent. For accurate calculation of the dynamical critical exponent is also necessary to calculate the values of the characteristic frequencies at higher linear dimensions. But our computational power were not allowed to carry out calculations for a system with a large number of spins. A rough estimate of the dependence (9) shows that the obtained value of the dynamical critical exponent for two values of exchange interactions ratio *R* is close to the value of *z*≈2. Note, that the value of index *z*=2 is corresponds to the theoretical predictions for anisotropic magnetic systems (model *A* in [11]).

## 5. CONCLUSIONS

Our results showed that the numerical investigation of the static and dynamic critical behavior of models of complex planar magnetic materials (such as iron-vanadium superlattice) is possible using the Hamiltonian of the Heisenberg model with the strong easy-plane anisotropy. Results of the investigation of the static critical behavior of superlattices at two values of interlayer and intralayer exchange interactions ratrio $R$ with the Hamiltonian (3) are in good agreement with results obtained for similar models using the Hamiltonian (1).

The obtained data show that the change of ratio between the intralayer and interlayer exchange interactions leads to a slight change in the shape of the dynamic structure factor and the change of its maximum value. This changes affect the dependence of the characteristic frequency $\omega_c$ on the linear dimensions of the system. Thus, changes in the values of $R$, according to (9), should lead to a change in the value of the dynamic critical exponent $z$. For accurate calculation of the dynamical critical exponent $z$ is necessary to investigate systems with a large linear dimensions.

This work was supported by the Federal Target Program "Research and scientific-pedagogical cadres of Innovative Russia" for 2009-2013 (State number P554 and № 02.740.11.03.97).

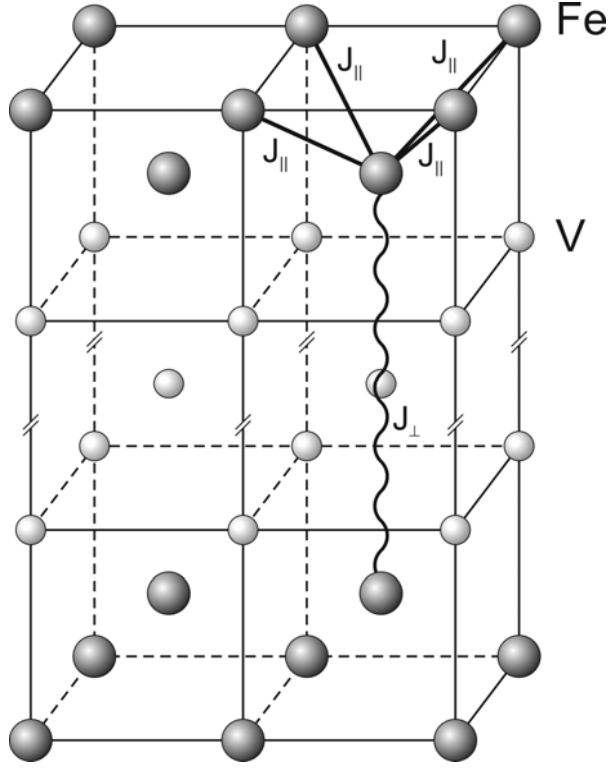

Fig.1. Schematic representation of iron-vanadium superlattice $Fe_2/V_{13}$. For clearness here are shown three vanadium monolayers of thirteen.

Table 1. Critical temperatures $T_c$ and critical exponents of magnetization β, susceptibility γ, and correlation lengt ν of $Fe_2/V_{13}$ superlattices at different values of ratio of the interlayer and intralayer exchange interactions $R$.

| $R$ | $T_c$ | β | γ | ν |
|---|---|---|---|---|
| *The results of the present work, Hamiltonian (3)* | | | | |
| 1.0 | 1.242(5) | 0.346(5) | 1.305(5) | 0.677(5) |
| 0.7 | 1.155(5) | 0.342(5) | 1.290(5) | 0.670(5) |
| *The results of works [6-8], Hamiltonian (1)* | | | | |
| 1.0 | 1.746(3) | 0.341(3) | 1.339(3) | 0.670(3) |
| 0.7 | 1.619(3) | 0.339(3) | 1.328(3) | 0.669(3) |
| *The classical 3d XY-model [24, 27]* | | | | |
| - | ~2.200 | 0.3485(2) | 1.3177(5) | 0.67155(27) |

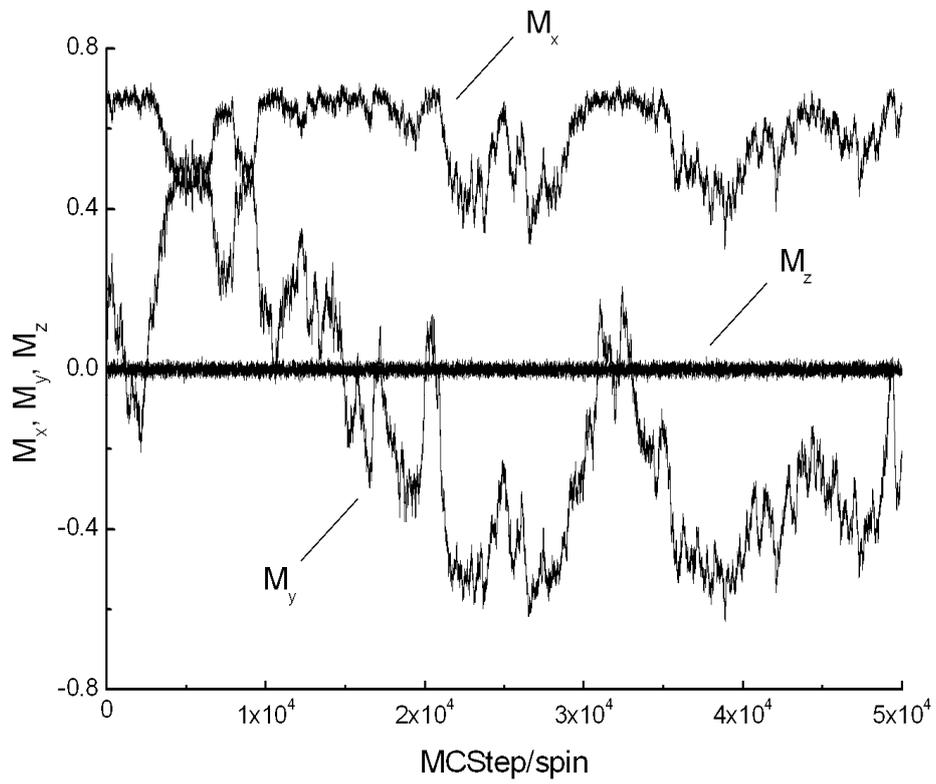

Fig. 2. Dependence of instantaneous values of vector magnetization components on the Monte-Carlo steps per spin. $r=1.0$, $N=2744$.

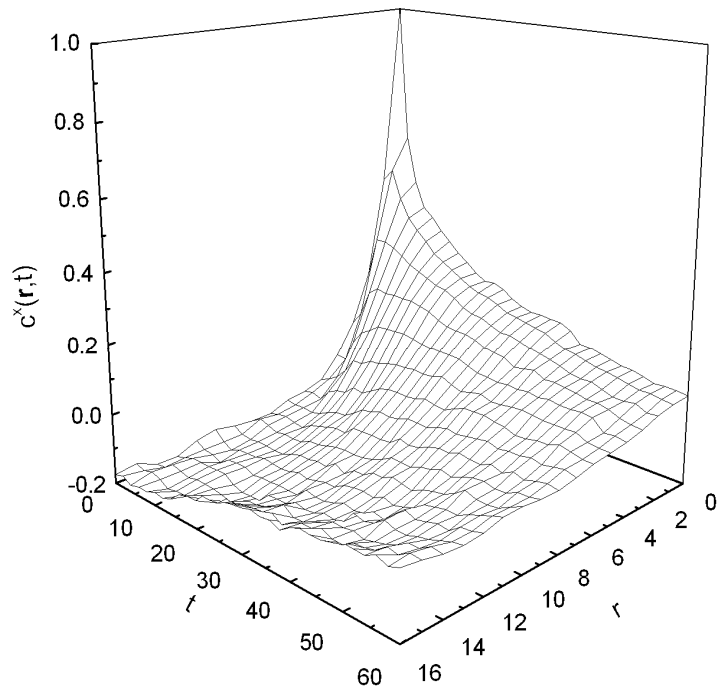

Fig. 3. Space-time spin correlation functions plotted by *x*-components of the spins in the line of *Ox*. $r=1.0$, $N=1728$, $T=T_c$.

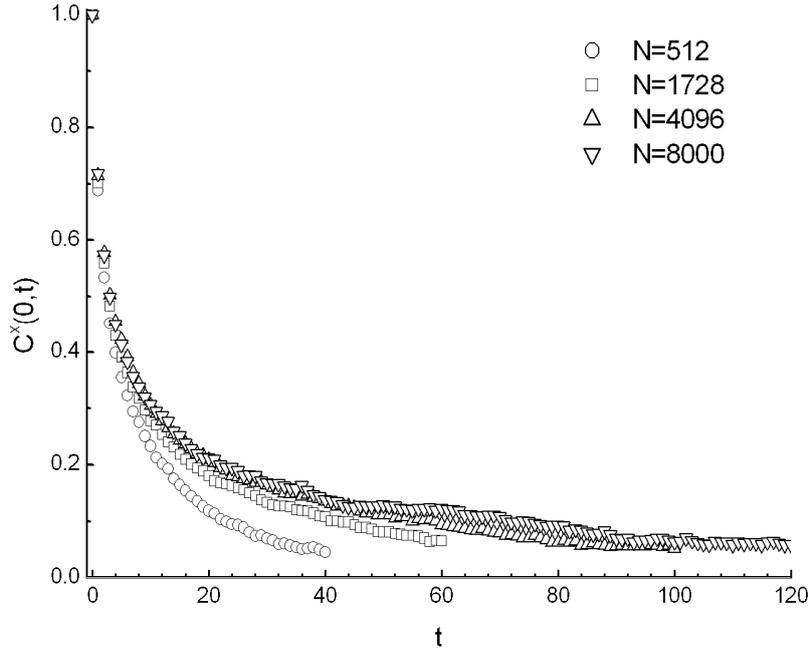

Fig. 4. Dependence of space-time spin correlation functions on the time at fixed $|\vec{r}| = 0$ for systems with different spin number $N$. Dependences are plotted by $x$-components of the spins in the line of $Ox$. $r=1.0$, $N=1728$, $T=T_c$.

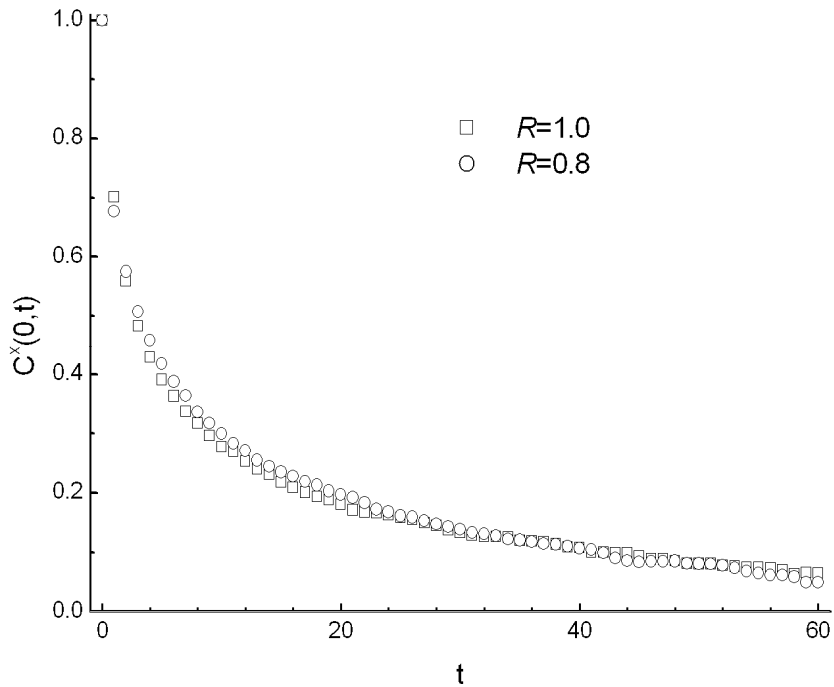

Fig. 5. Dependence of space-time spin correlation functions on the time at fixed $|\vec{r}| = 0$ for systems with different spin number N=1728 at two ratio of the interlayer and intra-layer exchange interactions $R$. Dependences are plotted by $x$-components of the spins in the line of $Ox$. $r=1.0$, $N=1728$, $T=T_c$.

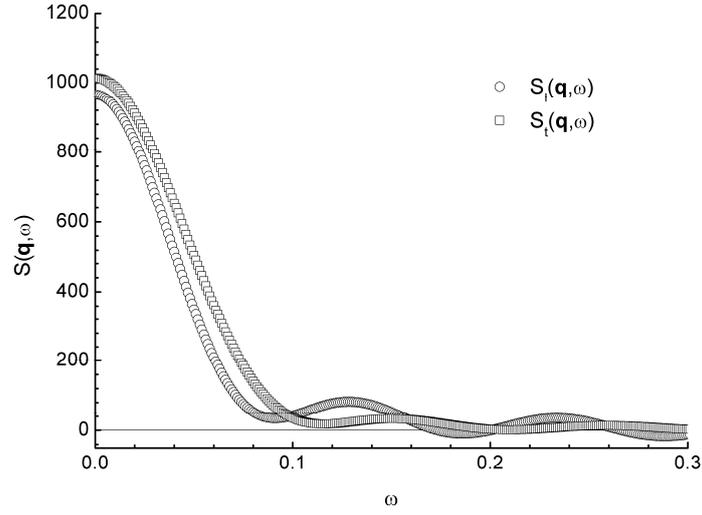

Fig. 6. Longitudinal $S_L(\vec{q},\omega)$ and transverse $S_T(\vec{q},\omega)$ dynamic structure factors for the systems with spin number $N=1728$ at $R=1.0$. $qL=2\pi$. The wave vector is directed along $Ox$ axes. $T=T_c$.

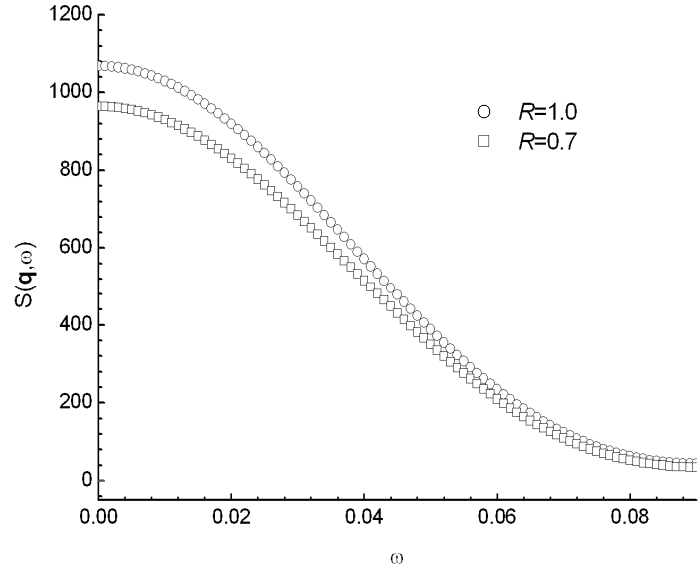

Fig. 7. Longitudinal $S_L(\vec{q},\omega)$ dynamic structure factor for the system with different spin number $N=1728$ at two ratio of exchange interactions $R$. $qL=2\pi$. The wave vector is directed along $Ox$ axes. $T=T_c$.

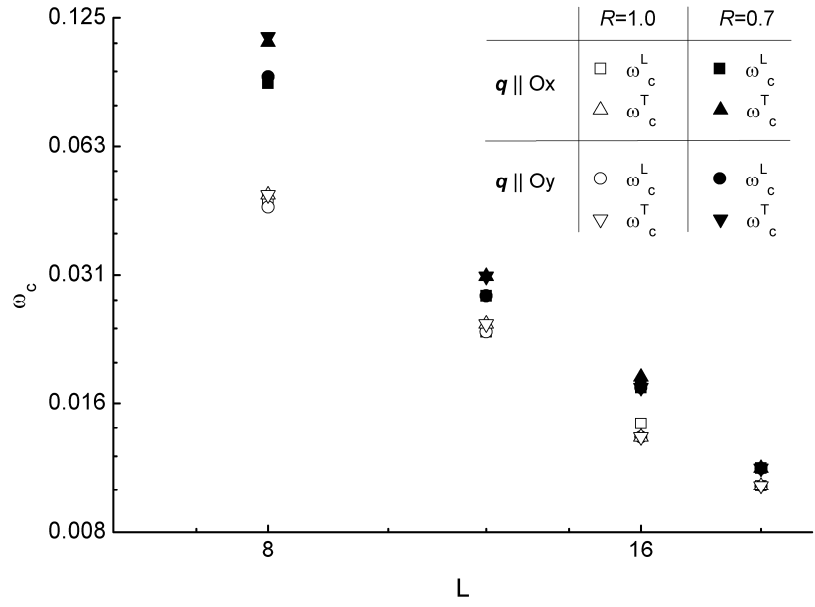

Fig. 8 Dependence of characteristic frequency $\omega_c$ on the linear sizes $L$ at two values of ratio of exchange interactions $R$. $T=T_c$.